\documentclass[twocolumn,superscriptaddress,showpacs,preprintnumbers,amsmath,amssymb]{revtex4}

\usepackage{graphicx}
\usepackage{dcolumn}
\usepackage{color}
\usepackage{bm}
\newcommand{\llangle}{\langle\!\langle}
\newcommand{\rrangle}{\rangle\!\rangle}
\newcommand{\upa}{\uparrow}
\newcommand{\dna}{\downarrow}

\newcommand{\beeq}{\begin{eqnarray}}
\newcommand{\eneq}{\end{eqnarray}}

\begin{document}


\title{Cooperon condensation and intra-valley pairing states in honeycomb Dirac systems}

\author{Shunji Tsuchiya}
\affiliation{Center for General Education, Tohoku Institute of Technology, 35-1 Yagiyamakasumi-cho, Taihaku-ku, Sendai 982-8577, Japan}
\email{tsuchiya@tohtech.ac.jp}

\author{Jun Goryo}
\affiliation{Department of Mathematics and Physics, Hirosaki University, Hirosaki 036-8561, Japan}

\author{Emiko Arahata}
\affiliation{Department of Physics, Tokyo Metropolitan University, 1-1 Minamiohsawa, Hachiohji, Tokyo 192-0397}

\author{Manfred Sigrist}
\affiliation{Institut f\"ur Theoretische Physik, ETH Z\"urich, CH-8093 Z\"urich, Switzerland}

\date{\today}

\begin{abstract}
Motivated by recent developments in the experimental study of superconducting graphene and transition metal dichalcogenides, we investigate superconductivity of the Kane-Mele (KM) model with short-range attractive interactions on the two-dimensional honeycomb lattice. We show that intra-valley spin-triplet pairing arises from nearest-neighbor (NN) attractive interaction and the intrinsic spin-orbit coupling. We demonstrate this in two independent approaches: We study superconducting instability driven by condensation of Cooperons, which are in-gap bound states of two conduction electrons, within the $T$-matrix approximation and also study the superconducting ground state within the mean-field theory. We find that Cooperons with antiparallel spins condense at the $K$ and $K'$ points. This leads to the emergence of an intra-valley spin-triplet pairing state belonging to the irreducible representation A$_1$ of the point group $C_{6v}$. The fact that this pairing state has opposite chirality for $K$ and $K'$ identifies this state as a ``helical" valley-triplet state, the valley-analog to the $^3$He-B phase in two dimension. Because of the finite center of mass momentum of Cooper pairs, the pair amplitude in NN bonds exhibits spatial modulation on the length scale of lattice constant, such that this pairing state may be viewed as a pair-density wave state. We find that the pair amplitude spontaneously breaks the translational symmetry and exhibits a $p$-Kekul\'e pattern. We also discuss the selection rule for pairing states focusing the characteristic band structure of the KM model and the Berry phase effects to the emergence of the intra-valley pairing state.
\end{abstract}

\maketitle

\section{Introduction}

Since the discovery of graphene, electronic properties of atomically thin two-dimensional (2D) materials have attracted wide-spread interest. Indeed remarkable features arise through the interplay of spin and valley degrees of freedom in the unusual band topology. 
Among other properties also superconductivity has been studied, despite great experimental difficulties in sample preparation and doping, particularly in graphene as well as transition metal dichalcogenides (TMDs). Superconductivity has been observed in Li-decorated monolayer graphene \cite{ludbrook-15}, ion gated MoSe$_2$, MoTe$_2$, WS$_2$ \cite{shi-15}, ion gated MoS$_2$ \cite{lu-15,saito-15}, and monolayer NbSe$_2$ \cite{xi-15}. In addition to their potential impact on applications, the superconducting states in such 2D materials also stimulate theoretical studies. Although the superconducting state observed in Li-decorated monolayer graphene is most likely due to conventional BCS-pairing arising from enhanced electron-phonon coupling by the adatoms \cite{profeta-12}, various exotic superconducting states have been suggested for pure and doped graphene \cite{uchoa-07,honerkamp-08,roy-10,nandkishore-12,kiesel-12,zhou-13,roy-14}. Furthermore, unconventional Ising pairing protected by spin-valley locking is predicted for the superconducting state in NbSe$_2$ atomic layers \cite{xi-15} and ion-gated MoS$_2$ \cite{lu-15,saito-15}.  
\par
Motivated by these experimental advances, we investigate superconductivity in the 2D honeycomb lattice structure that is common to graphene and TMDs.
Our main purpose in this paper is to analyze the structure of the superconducting phase in the honeycomb lattice with special emphasis on topological aspects. For this purpose, we employ the Kane-Mele (KM) model \cite{kane-05} that was proposed as a minimal model of topological insulators \cite{hasan-10,qi-11}. We assume generic short-range attractive interactions and discuss the symmetry of superconducting ground states. In contrast to most studies on superconductivity our starting point will be the insulating state where we explore the pairing states that could arise through Cooperon condensation for sufficiently strong pairing interactions. As we will discuss below a particularly interesting case of unconventional Cooper pairing appears for nearest-neighbour (NN) attractive interaction.
\par
\begin{figure}
\centering
\includegraphics[width=\linewidth]{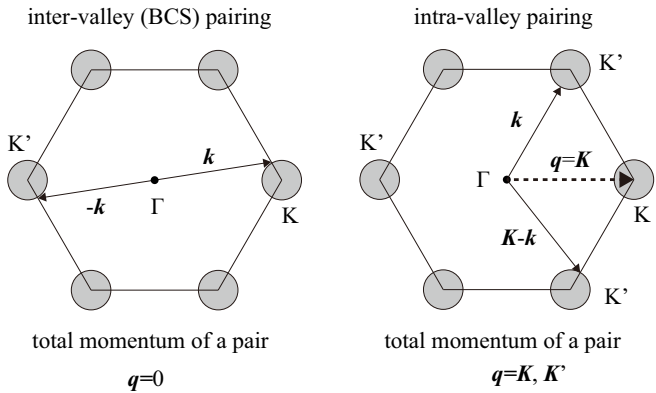}
\caption{Schematic illustration of the inter-valley and intra-valley pairing states. The gray circles represent the Fermi circles. In the former, the center of mass momentum of a pair is zero. In the latter, the pair has finite center of mass momentum $\bm q=\bm K$ or $\bm K'$.}
\label{fig:intravalley}
\end{figure}

The two possible pairing states on the honeycomb lattice considering the valley-structure of the electronic bands are illustrated in Fig.~\ref{fig:intravalley}: {\it Inter-valley} pairing state and {\it intra-valley} pairing state. The former is the simple BCS pairing state involving electrons with opposite momenta in the different valleys near the $K$ and $K'$ points. In contrast, electrons form pairs within the same valley in the latter case. Namely, they have opposite momenta with respect to $K$ or $K'$ points, and, therefore, an electron pair has finite center of mass momentum equivalent to ${\bm K}'$ and $\bm K$, respectively. Because of the finite center of mass momentum of Cooper pairs, this pairing state may be viewed as a pair-density wave (PDW) state \cite{yoshida-12}, in which the gap function spatially modulates on length scales of the lattice constant. The possibility of the intra-valley pairing has been pointed out in graphene \cite{roy-10,zhou-13} as well as in doped Weyl semimetals \cite{cho-12}.
\par
In this paper, we show that the {\it intra-valley spin-triplet} pairing state can arise due to the interplay of the NN attractive interaction and the intrinsic spin-orbit (SO) coupling in the KM model. The interesting feature of the intra-valley pairing state is that it involves two gap functions associated with Cooper pairs condensed at each of the two valleys, $K$ and $K'$ point (see Fig.~\ref{fig:intravalley}). In the intra-valley spin-triplet pairing state, the gap functions have both the components of $s$ and $p$-wave symmetry in the vicinity of $\bm K$ and $\bm K'$, and constitute a parity-mixed superconducting state, as we will show. 
We demonstrate the emergence of this exotic superconducting state by employing two independent microscopic approaches: We first study superconducting instability in the insulating state within the $T$-matrix approximation, and then we examine the most stable superconducting state within the mean-field (MF) theory. In the former, we find that bound states of two conduction electrons called ``Cooperons" \cite{kohmoto-90,nozieres-99,rice-12,tsuchiya-13} are formed within the band gap and the intra-valley pairing state is preempted by condensation of Cooperons at the $K$ and $K'$ points at the same interaction strength. We also discuss the origin and nature of the intra-valley pairing state. We find that it may arise due to the Berry phase effects associated with the Dirac points, i.e., $K$ and $K'$ points.
\par
The paper is organized as follows: In Sec.~\ref{sec.model}, we describe the system and the model. In Sec.~\ref{sec.rule}, we discuss the selection rule for pairing states based on the characteristic feature of the energy band. In Sec.~\ref{sec.Cooperon}, we study formation of Cooperons and their condensation in the topological insulating state. In Sec.~\ref{sec.mftheory}, we study the superconducting ground state within the MF theory and discuss its various aspects. We conclude in Sec.~\ref{sec.conclusions}.

\section{Model}
\label{sec.model}
\begin{figure}
\centerline{\includegraphics[width=4cm]{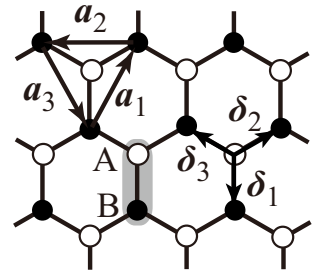}}
\caption{The honeycomb lattice with the basis vectors. The unit cell (gray region) consists of the sublattices A and B. $\bm \delta_i$ and $\bm a_i$ ($i=1,2,3$) are the bond vectors between NN and NNN sites, respectively. We set the lattice constant unity ($|\bm a_i|=1$).}
\label{fig.honeycomblattice}
\end{figure}
We study the KM model \cite{kane-05} with short-range attractive interaction on the honeycomb lattice depicted in Fig.~\ref{fig.honeycomblattice}. The Hamiltonian reads
\begin{eqnarray}
H&=&H_{\rm KM}+H_{\rm int},\label{eq.Hamiltonian}\\
H_{\rm KM}&=&-t\sum_{\langle
 i,j\rangle}\sum_\sigma \left(c_{i\sigma}^\dagger c_{j\sigma}+{\rm h.c.}\right)-\mu\sum_{i,\sigma}n_{i\sigma}\nonumber\\
&&-it'\sum_{\llangle
  i,j\rrangle}\sum_{\sigma,\sigma^\prime}\nu_{ij}(\sigma_z)_{\sigma \sigma^\prime} c_{i\sigma}^\dagger c_{j\sigma^\prime},\label{eq.KaneMele}\\
H_{\rm int}&=&-U\sum_i n_{i\uparrow}n_{i\downarrow}-V\sum_{\langle
  i,j\rangle} n_{i}n_{j},
\label{eq.interaction}
\end{eqnarray}
where $c_{i\sigma}$ annihilates an electron at site $i$ with spin $\sigma$, $\mu$ the chemical potential, and $\langle i,j\rangle / \llangle i,j\rrangle$ denotes the summation over all the NN/next-nearest-neighbor (NNN) sites. The first term in Eq.~(\ref{eq.KaneMele}) describes the NN hopping and the third term the intrinsic SO coupling \cite{kane-05}, where $\sigma_\rho$ ($\rho=x,y,z$) is the Pauli matrix of electron spin and $\nu_{ij}=1$ (-1) if electrons make a left (right) turn to get to the site $i$ from the site $j$. We consider the on-site and NN attractive interactions in Eq.~(\ref{eq.interaction}) and assume $U, V>0$.
\par
Turning to $ {\bm k} $-space, we introduce
\begin{equation}
c_{i\sigma}=\frac{1}{\sqrt{M}}\sum_{\bm k}c_{\bm k\sigma}e^{-i\bm k\cdot \bm r_i}\ ,
\end{equation}
where $M=N/2$ is the total number of unit cells that is half of the total lattice sites $N$. The KM Hamiltonian (\ref{eq.KaneMele}) in momentum space reads
\begin{equation}
H_{\rm KM}=\sum_{\bm k}\left[\psi_{\bm k\uparrow}^\dagger
\left(
\begin{array}{cc}
\zeta_{\bm k} & \gamma_{\bm k} \\
\gamma_{\bm k}^* & -\zeta_{\bm k}
\end{array}
\right)
\psi_{\bm k\uparrow}
+
\psi_{\bm k\downarrow}^\dagger
\left(
\begin{array}{cc}
-\zeta_{\bm k} & \gamma_{\bm k} \\
\gamma_{\bm k}^* & \zeta_{\bm k}
\end{array}
\right)
\psi_{\bm k\downarrow}\right],
\label{eq.KMkspace}
\end{equation}
where,  
$\psi_{\bm k\sigma}=(a_{\bm k\sigma}, b_{\bm k\sigma})^T$, 
$\gamma_{\bm k}=-t(e^{-i\bm k\cdot {\bm \delta}_1}+e^{-i\bm k\cdot {\bm \delta}_2}+e^{-i\bm k\cdot {\bm \delta}_3})$, and $\zeta_{\bm k}=2t'(\sin\bm k\cdot\bm a_1+\sin\bm k\cdot\bm a_2+\sin\bm k\cdot\bm a_3)-\mu$. 
Here, $a_{\bm k\sigma}$ ($b_{\bm k\sigma}$) annihilates an electron on the A (B) sublattice with momentum $\bm k$ and spin $\sigma$. ${\bm \delta}_i$ and $\bm a_i$ ($i=1,2,3$) are the bond vectors that connect the NN sites and NNN sites, respectively, as shown in Fig.~\ref{fig.honeycomblattice}. 
We set the lattice constant unity ($|{\bm a}_i|=1$).
\par
The dispersion relations of the conduction and valence bands are obtained by diagonalizing Eq.~(\ref{eq.KMkspace}) as 
\begin{equation}
E=\pm\sqrt{|\gamma_{\bm k}|^2+\zeta_{\bm k}^2}=\pm\epsilon_{\bm k}.
\end{equation}
$\gamma_{\bm k}$ is approximated in the vicinity of the $K$ point $(4\pi/3,0)= \bm K$ as 
\beeq
\gamma_{\bm K+\bm p}\simeq v_F (p_x-i p_y),
\label{aroundK}
\eneq
and the $K'$ point $(-4\pi/3,0)=\bm K'=-\bm K$ as 
\beeq
\gamma_{\bm K'+\bm p}=\gamma_{\bm K - \bm p}^*\simeq -v_F  (p_x+i p_y), 
\label{aroundKp}
\eneq
where $\bm p$ denotes momentum measured relative to the $K$ and $K'$ points ($\bm p=\bm k-\bm K$, $\bm k-\bm K'$, $p\ll |\bm K|$).
 Here, we introduced the Fermi velocity $v_F=\sqrt{3}t/2$. Thus, at half-filling ($\mu=0$) without the SO coupling ($t'=0$), the conduction and valence bands have linear dispersions $\epsilon_{\bm k}=|\gamma_{\bm k}|=v_Fp$ that describe massless Dirac fermions in the vicinity of the $K$ and $K'$ points. 
\par
\begin{figure}
\centerline{\includegraphics[width=8cm]{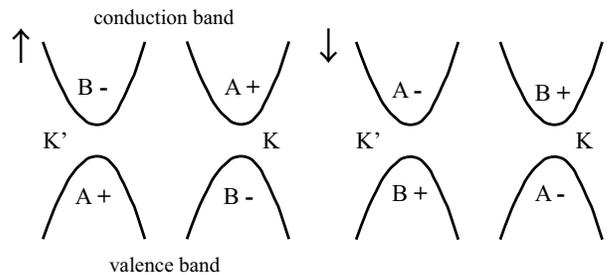}}
\caption{Schematic illustration of the energy band of Dirac fermions described in the KM model (\ref{eq.linearKM}). The ``A" or ``B" assigned to each valley means the sublattice at which the wave function of the bottom of the valence band or top of the conduction band localizes. The symbols ``$+$" and ``$-$" denote the sign of the Berry phase associated with adiabatic evolution within the energy band around the $K$ or $K'$ point.}
\label{energyband}
\end{figure}

On the other hand, the diagonal elements in Eq.~(\ref{eq.KMkspace}) are approximated as
\beeq
\zeta_{\bm K+\bm p}\simeq \Delta_{\rm SO}, \quad\zeta_{\bm K'+\bm p}\simeq -\Delta_{\rm SO},
\eneq
where $\Delta_{\rm SO}= 3\sqrt{3}t'$ (we assume $t'>0$ throughout the paper). The dispersion in the vicinity of the $K$ and $K'$ points at half-filling is given by 
\beeq
E=\pm\sqrt{v_F^2p^2+\Delta_{\rm SO}^2}~.
\label{eq.KMspctrm}
\eneq
Figure~\ref{energyband} schematically shows the dispersion (\ref{eq.KMspctrm}) that has the energy gap $2\Delta_{\rm SO}$ at the $K$ and $K'$ points. Thus, the low-energy physics is dominated by massive Dirac fermions.
\par
The effective Hamiltonian at half-filling linearized in the vicinity of the $K$ and $K'$ points reads
\beeq
&&H_{\rm KM}=\displaystyle\sum_{\bm p}\psi_{K\bm p}^\dagger
\left(v_F\bm \tau\cdot\bm p+\Delta_{\rm SO}\sigma_z \tau_z\right)\psi_{K\bm p}\nonumber\\
&&+\displaystyle\sum_{\bm p}\psi_{K'\bm p}^\dagger
\left(-v_F\bm \tau^\ast\cdot\bm p-\Delta_{\rm SO}\sigma_z \tau_z\right)\psi_{K'\bm p},
\label{eq.linearKM}
\eneq
where $\psi_{K\bm p}=(\psi_{\bm K+\bm p\upa},\psi_{\bm K+\bm p\dna})$ and $\tau_\rho$ ($\rho=x,y,z$) is the Pauli matrix of sublattice-pseudospin.
Precisely at the $K$ or $K'$ point, since the off-diagonal terms vanish, Eq.~(\ref{eq.linearKM}) is diagonalized in the sublattice basis. This means that the wave functions at the bottom of the conduction band and the top of the valence band localize on either A or B sublattice. Figure~\ref{energyband} shows the sublattices assigned to them. It exhibits a peculiar character of the wave function in momentum space: The sublattices assigned to the $K$ and $K'$ points are different within the same band. This implies that the insulating state due to the SO coupling described by the KM Hamiltonian (\ref{eq.KaneMele}) does not reduce to the trivial band insulator with decoupled A and B sublattices in the limit of large energy gap $\Delta_{\rm SO}\gg t$. Thus, it is topologically distinct from the trivial band insulator \cite{kane-05}. 
In this peculiar insulating state, the spin Hall conductivity is quantized, which is characterized by the topological number called spin Chern number. The nonzero spin Chern number guarantees the existence of the helical edge modes that are predicted by the bulk/boundary correspondence \cite{kane-05,hasan-10,qi-11}.
\par
Note that the Berry phase of Bloch electrons associated with adiabatic evolution around the $K$ and $K'$ points in momentum space has opposite signs. In particular, for a massless Dirac fermion ($t'=0$), the Berry phase of conduction band upon going around the $K$ and $K'$ points are $\pi$ and $-\pi$, respectively. This feature plays a crucial role in the emergence of the intra-valley pairing state, as we will discuss in Sec.~\ref{sec.Cooperon}. 

\section{Selection rule for pairing states}
\label{sec.rule}

The special character of the wave function of the KM model described in the last section enables us to identify possible pairing states induced by the local attractive interactions which we choose to be of density-density type to avoid any bias on the spin configuration. On the other hand, through the choice of sublattices we select at the outset different sublattice pseudo-spin configurations. 
\par
Figure~\ref{energyband} implies that in the inter-valley pairing state two conduction electrons in different valleys form a pair. With the on-site attractive interaction electrons pair on the same sublattice with opposite spins, while the NN interaction couples electrons on different sublattices and favors pairing with parallel spins. On the other hand, in the intra-valley pairing state the NN interaction prefers opposite spins. The same applies to two holes in the valence band.
\par
We can extend the above observation further to more general attractive interactions to derive the following selection rule: 
If the attractive interaction dominantly works between electrons (holes) on the same sublattice, it induces {\it inter-valley} pairing of electrons (holes) with {\it opposite} spins or {\it intra-valley} pairing with {\it parallel} spins. If the attractive interaction dominantly works between electrons (holes) on different sublattices, it causes {\it inter-valley} pairing of electrons (holes) with {\it parallel} spins or {\it intra-valley} pairing with {\it opposite} spins.
\par
Indeed, the on-site attractive interaction naturally induces the inter-valley pairing, i.e., the conventional spin-singlet $s$-wave BCS pairing. In contrast, the NN attractive interaction induces the unconventional intra-valley pairing state with mixed parity, as we will see in the next section.

\section{Cooperon condensation}
\label{sec.Cooperon}

  In an insulator, superconducting fluctuation due to attractive interaction leads to formation of Cooperons within the band gap and a superconducting instability could be driven by condensation of Cooperons \cite{kohmoto-90,nozieres-99,rice-12,tsuchiya-13}. In this section,  to verify the selection rule of the previous section from a microscopic approach, we study formation and condensation of Cooperons in the topological insulating state at half-filling based on the tight-binding Hamiltonian (\ref{eq.Hamiltonian}).
\par
The Green's function in a matrix form in the sublattice-pseudospin space is given by
\beeq
\hat G_\sigma(\bm k, \tilde t-\tilde t')=-\langle T_{\tilde t} \psi_{\bm k\sigma}(\tilde t)\psi_{\bm k\sigma}^\dagger(\tilde t') \rangle,
\eneq
where $\tilde t$ denotes imaginary time. The Green's function for spin-up electrons in momentum space reads
\begin{eqnarray}
\hat G_{\uparrow}(k)&=&\frac{1}{i\omega_n-\left(\begin{array}{cc}\zeta_{\bm k} & \gamma_{\bm k} \\ \gamma_{\bm k}^* & -\zeta_{\bm k} \end{array} \right)+\mu}\nonumber\\
&=&\frac{ \hat P_{\bm k\uparrow}}{i\omega_n-\epsilon_{\bm k}+\mu}+
\frac{ \hat Q_{\bm k\uparrow}}{i\omega_n+\epsilon_{\bm k}+\mu},\label{eq.Gup}\\
\hat P_{\bm k\uparrow}
&=&
\left(
\begin{array}{cc}
u_{\bm k}^2 & u_{\bm k}v_{\bm k} e^{i\theta_{\bm k}}\\
u_{\bm k} v_{\bm k} e^{-i\theta_{\bm k}} & v_{\bm k}^2
\end{array}
\right),\\
\hat Q_{\bm k\uparrow}&=&
\left(
\begin{array}{cc}
v_{\bm k}^2 & -u_{\bm k}v_{\bm k} e^{i\theta_{\bm k}}\\
-u_{\bm k} v_{\bm k} e^{-i\theta_{\bm k}} & u_{\bm k}^2
\end{array}
\right), 
\end{eqnarray}
where $\omega_n$ is the fermionic Matsubara frequency and $e^{i\theta_{\bm k}}=\gamma_{\bm k}/|\gamma_{\bm k}|$. $u_{\bm k}$ and $v_{\bm k}$ are defined as
\beeq
u_{\bm k}&=&\sqrt{\frac{1}{2}\left(1+\frac{\zeta_{\bm k}}{\epsilon_{\bm k}}\right)},
v_{\bm k}=\sqrt{\frac{1}{2}\left(1-\frac{\zeta_{\bm k}}{\epsilon_{\bm k}}\right)}.
\label{eq.ukvk}
\eneq
The Green's function for spin-down electrons can be obtained by
substituting $\zeta_{\bm k}\to -\zeta_{\bm k}$ in $\hat G_{\uparrow}$ as
\begin{eqnarray}
\hat G_{\downarrow}(k)&=&\frac{1}{i\omega_n-\left(\begin{array}{cc}-\zeta_{\bm k} & \gamma_{\bm k} \\ \gamma_{\bm k}^* & \zeta_{\bm k} \end{array} \right)+\mu}\nonumber\\
&=&\frac{ \hat P_{\bm k\downarrow}}{i\omega_n-\epsilon_{\bm k}+\mu}+
\frac{ \hat Q_{\bm k\downarrow}}{i\omega_n+\epsilon_{\bm k}+\mu},\label{eq.Gdown}\\
\hat P_{\bm k\downarrow}&=&
\left(
\begin{array}{cc}
v_{\bm k}^2 & u_{\bm k}v_{\bm k} e^{i\theta_{\bm k}}\\
u_{\bm k} v_{\bm k} e^{-i\theta_{\bm k}} & u_{\bm k}^2
\end{array}
\right),\\ 
\hat Q_{\bm k\downarrow}&=&
\left(
\begin{array}{cc}
u_{\bm k}^2 & -u_{\bm k}v_{\bm k} e^{i\theta_{\bm k}}\\
-u_{\bm k} v_{\bm k} e^{-i\theta_{\bm k}} & v_{\bm k}^2
\end{array}
\right).
\end{eqnarray}
Note that the phase factor in the off-diagonal elements is associated with the flip of the sublattice-pseudospin.
\par
The interaction Hamiltonian (\ref{eq.interaction}) in momentum space reads
\beeq
H_{\rm int}
&=&\frac{1}{2M}\sum_{\bm k,\bm k',\bm q}\sum_{\sigma,\sigma'}\sum_{\tau_,\tau'}g^{\tau\tau'}_{\sigma\sigma'}(\bm k'-\bm k)\nonumber\\
&&\times c^\dagger_{\bm k\tau\sigma}c^\dagger_{-\bm k+\bm q\tau'\sigma'}c_{-\bm
 k'+\bm q\tau'\sigma'}c_{\bm k'\tau\sigma}~,\\
 g^{\tau\tau'}_{\sigma\sigma'}(\bm k)&=&-U\delta_{\sigma',\bar\sigma}\delta_{\tau,\tau'}\nonumber\\
 &&-V\left[\delta_{\tau A}\delta_{\tau'B}f^*(\bm k)+\delta_{\tau B}\delta_{\tau'A}f(\bm k)\right],
\eneq
where {$c_{\bm k\tau\sigma}$ annihilates an electron with momentum $\bm k$ and spin $\sigma$ at sublattice $\tau$, $\bar\sigma$ denotes opposite spin of $\sigma$, and} $f(\bm k)=\gamma_{\bm k}/(-t)$.
\par
\begin{figure}
\centering
\includegraphics[width=\linewidth]{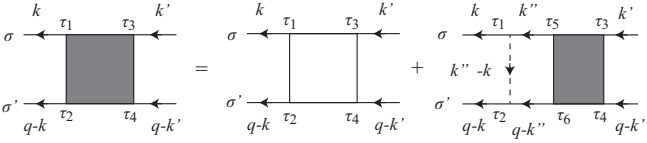}
\caption{Diagrammatic representation of the BS equation for the $T$-matrix approximation. The gray regions represent the vertex part $\Gamma$.}
\label{diagram}
\end{figure}

We employ the $T$-matrix approximation that describes the superconducting instability due to pair formation. The Bethe-Salpeter (BS) equation for the $T$-matrix approximation diagrammatically represented in Fig.~\ref{diagram} is given by
\begin{eqnarray}
&& \Gamma_{\sigma\sigma'}^{\tau_1\tau_2,\tau_3\tau_4}(\bm k,\bm k';q)=\Gamma_{\sigma\sigma'}^{0\tau_1\tau_2,\tau_3\tau_4}(\bm k,\bm k')\nonumber\\
&&-\frac{1}{\beta M}\sum_{\bm k'',\omega_n''}\sum_{\nu_5,\nu_6}g^{\tau_1\tau_2}_{\sigma\sigma'}({\bm k}''-{\bm k}')G_\sigma^{\tau_1\tau_5}(k'')\nonumber\\
&&\times G^{\tau_2\tau_6}_{\sigma'}(q-k'')\Gamma^{\tau_5\tau_6,\tau_3\tau_4}_{\sigma\sigma'}({\bm k}'',{\bm k}';q),
\label{eq.BS}
\end{eqnarray}
where $\Gamma$ is the vertex part. In lowest-order, it reduces to the bare interaction:
\begin{eqnarray}
\Gamma^{0\tau_1\tau_2,\tau_3\tau_4}_{\sigma\sigma'}(\bm k,\bm k')=\delta_{\tau_1,\tau_3}\delta_{\tau_2,\tau_4}g^{\tau_1\tau_2}_{\sigma\sigma'}(\bm k'-\bm k).
\end{eqnarray}
We denote $k=(\bm k,i\omega_n)$ and $q=(\bm q,i\Omega_n)$, where $\Omega_n$ is the bosonic Matsubara frequency. Hereafter in this section, we restrict ourselves within the insulating state at half-filling and set $\mu=0$.

\subsection{On-site attractive interaction}

We first set $V=0$ to examine pairing due to the on-site attractive interaction. In this case, Eq.~(\ref{eq.BS}) greatly simplifies to 
\beeq
&&\hat\Gamma(q)=-U\hat I+U\hat\Pi(q)\hat\Gamma(q),\label{eq.BS_onsite}\\
&&\Pi^{\tau_1\tau_2}(q)=\frac{1}{\beta M}\sum_{\bm k,\omega_n}G^{\tau_1\tau_2}_\sigma(k) G^{\tau_1\tau_2}_{\bar\sigma}(q-k),\label{eq.Pi_onsite}
\eneq
where $\Gamma_{\sigma\sigma'}^{\tau_1\tau_2,\tau_3\tau_4}(\bm k,\bm k';q)=\delta_{\tau_1,\tau_2}\delta_{\tau_3,\tau_4}\delta_{\sigma',\bar\sigma}\Gamma^{\tau_1\tau_3}(q)$, $(\hat \Gamma)_{\tau_1\tau_2}=\Gamma^{\tau_1\tau_2}$, and $(\hat \Pi)_{\tau_1\tau_2}=\Pi^{\tau_1\tau_2}$. Eq.~(\ref{eq.BS_onsite}) is easily solved:
\beeq
\hat\Gamma(q)=(-U)\left(\hat I-U\hat\Pi(q)\right)^{-1}.
\eneq
\par
From the condition for $\hat\Gamma(q)$ to have poles,
\beeq
{\rm det}[\hat I-U\hat\Pi(\bm q,\Omega)]=0,
\label{eq.cooperonU}
\eneq
we obtain the energy spectrum of Cooperons.
\begin{figure}
\centering
\includegraphics[width=\linewidth]{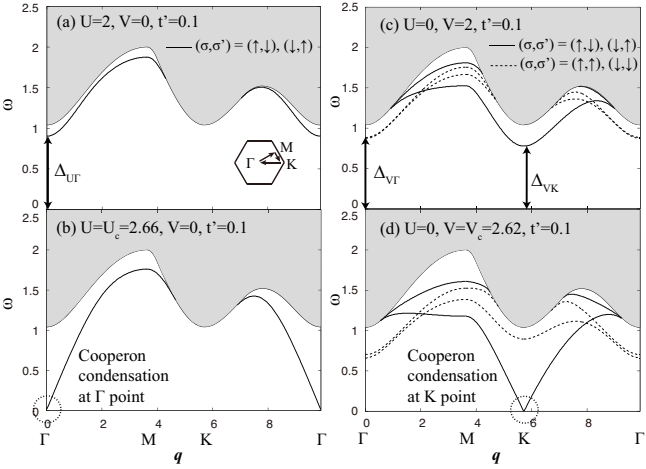}
\caption{Energy spectrum of Cooperons formed by the on-site attractive interaction ((a) and (b)) and NN attractive interaction ((c) and (d)) with $t'=0.1$ in all cases. Energies are in units of $t$. The solid (dashed) lines show the dispersion of a Cooperon composed of electrons with opposite (parallel) spins. The gray region represents the two-particle continuum. Cooperons have the minimum energy at the $\Gamma$ point in (a) and (b), which is denoted by $\Delta_{\rm U\Gamma}$, while they have the minimum energy at the $K$ point in (c) and (d), which is denoted by $\Delta_{\rm VK}$.}
\label{fig.cooperon}
\end{figure}
\begin{figure}
\centering
\includegraphics[width=7cm]{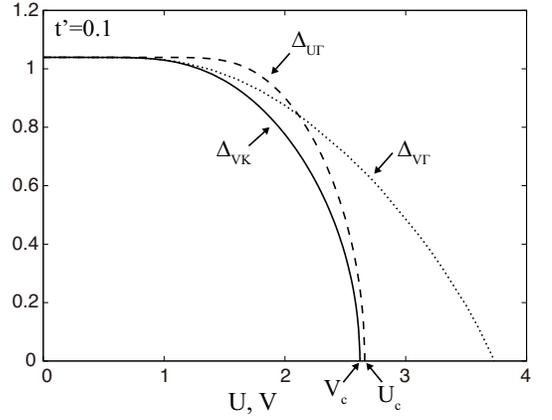}
\caption{Energy gap of Cooperons by the on-site attraction at the $\Gamma$ point ($\Delta_{\rm U\Gamma}$) as a function of $U$, and those of Cooperons by the NN attraction at the $K$ point ($\Delta_{\rm VK}$) and at the $\Gamma$ point ($\Delta_{\rm V\Gamma}$) for $t'=0.1$ (see Fig.~\ref{fig.cooperon}). Energies are in units of $t$. Cooperons soften at the critical strength $U_{\rm c}$ and $V_{\rm c}$.  }
\label{fig.cooperongap}
\end{figure}
\par
Figures~\ref{fig.cooperon} (a) and (b) show the energy spectrum of Cooperons obtained by solving Eq.~(\ref{eq.cooperonU}). They illustrate the formation of Cooperons below the edge of the two-particle continuum. Any small $U>0$ induces Cooperons below the continuum. The on-site attractive interaction boosts the formation of a Cooperon bound state, particularly, in the vicinity of the $\Gamma$ point at which the dispersion has its minimum. This implies that the inter-valley pairing of two electrons is energetically favorable.
\par
The minimum energy gap at the $\Gamma$ point ($\Delta_{\rm U\Gamma}$) is plotted as a function of $U$ in Fig.~\ref{fig.cooperongap}. $\Delta_{\rm U\Gamma}$ progressively decreases as $U$ is increased and the Cooperon softens and eventually reaches zero energy at the $\Gamma$ point for the critical strength $U_{\rm c}$, as shown in Fig.~\ref{fig.cooperon} (b) indicating an instability. The condensation of Cooperons at the $\Gamma$ point leads to the proliferation of Cooper pairs with zero total momentum, i.e., the inter-valley pairing state. Thus, the conventional $s$-wave spin-singlet superconducting state is realized due to the on-site attractive interaction.

\subsection{NN attractive interaction}

We next set $U=0$ and examine pairing due to the NN attractive interaction. Since $\Gamma^{\tau_1\tau_2,\tau_3\tau_4}$ vanishes if $\tau_1=\tau_2$ or $\tau_3=\tau_4$, the nonzero matrix elements of $\Gamma^{\tau_1\tau_2,\tau_3\tau_4}$ are those with $(\tau_1,\tau_2;\tau_3,\tau_4)=(A,B;A,B)$, $(A,B;B,A)$, $(B,A;A,B)$, and $(B,A;B,A)$. Thus, Eq.~(\ref{eq.BS}) can be rewritten in a matrix form as
\begin{eqnarray}
&&\hat \Gamma_{\sigma\sigma'}(\bm k,\bm k'; q)=\hat \Gamma^0(\bm k,\bm k')\nonumber\\
&&-\frac{1}{M}\sum_{\bm k''}\hat \Gamma^0(\bm k,\bm k'')\hat \pi_{\sigma\sigma'}(\bm k'';q)\hat\Gamma_{\sigma\sigma'}(\bm k'',\bm k';q).
\label{eq.BS_NN}
\end{eqnarray}
Here, we define
\beeq
&& \hat\Gamma_{\sigma\sigma'}(\bm k,\bm k';q)=
\left(
\begin{array}{cc}
\Gamma^{AB,AB}_{\sigma\sigma'}(\bm k,\bm k';q) & \Gamma^{AB,BA}_{\sigma\sigma'}(\bm k,\bm k';q)\\
\Gamma^{BA,AB}_{\sigma\sigma'}(\bm k,\bm k';q) & \Gamma^{BA,BA}_{\sigma\sigma'}(\bm k,\bm k';q)
\end{array}
\right),\\
&& \hat \pi_{\sigma\sigma'}(\bm k;q)\nonumber\\
&&=\frac{1}{\beta}\sum_{\omega_n}
\left(
\begin{array}{cc}
G_{\sigma}^{AA}(p)G_{\sigma'}^{BB}(q-k) & G_{\sigma}^{AB}(p)G_{\sigma'}^{BA}(q-k)\\
G_{\sigma}^{BA}(p)G_{\sigma'}^{AB}(q-k) & G_{\sigma}^{BB}(p)G_{\sigma'}^{AA}(q-k)
\end{array}
\right),\\
&&\hat\Gamma^0(\bm k,\bm k')=-V\sum_{i=1}^3\hat m_{\bm k}^i \hat m_{\bm k'}^{i\dag},
\eneq
where
\beeq
\hat m_{\bm k}^i=
\left(
\begin{array}{cc}
e^{-i\bm k\cdot \bm \delta_i} & 0 \\
0 & e^{i\bm k\cdot \bm \delta_i}
\end{array}
\right).
\eneq
We then obtain
\beeq
\hat X^i_{\sigma\sigma'}(\bm k;q)=\hat X^{0i}_{\sigma\sigma'}(\bm k;q)+V\sum_{j=1}^3 \hat \Pi_{\sigma\sigma'}^{ij}(q)\hat X^j_{\sigma\sigma'}(\bm k;q),
\label{eq.BS_NN2}
\eneq
where
\beeq 
&&\hat X^i_{\sigma\sigma'}(\bm k;q)=\frac{1}{M}\sum_{\bm k'} \hat m_{\bm k'}^{i\dag} \hat \pi_{\sigma\sigma'}(\bm k';q)\hat\Gamma_{\sigma\sigma'}(\bm k',\bm k;q),\label{eq.X}\\
&&\hat X^{0i}_{\sigma\sigma'}(\bm k;q)=\frac{1}{M}\sum_{\bm k'} \hat m_{\bm k'}^{i\dag} \hat \pi_{\sigma\sigma'}(\bm k';q)\hat\Gamma^0(\bm k',\bm k),\label{eq.X0}\\
&&\hat\Pi_{\sigma\sigma'}^{ij}(q)=\frac{1}{M}\sum_{\bm k}\hat m_{\bm k}^{i\dag}\hat\pi_{\sigma\sigma'}(\bm k;q)\hat m_{\bm k}^j.
\eneq
Eq.~(\ref{eq.BS_NN2}) can be further cast into the following form
\begin{eqnarray}
&& \tilde X_{\sigma\sigma'}(\bm k;q)=\tilde X_{\sigma\sigma'}^0(\bm k;q)+V\tilde\Pi_{\sigma\sigma'}(q)\tilde X_{\sigma\sigma'}(\bm k;q),\label{eq.BS_NN3}\\
&&\tilde X_{\sigma\sigma'}(\bm k;q)=\left(
\begin{array}{c}
\hat X^1_{\sigma\sigma'}(\bm k;q)\\
\hat X^2_{\sigma\sigma'}(\bm k;q)\\
\hat X^3_{\sigma\sigma'}(\bm k;q)
\end{array}
\right),\\
&&\tilde X^0_{\sigma\sigma'}(\bm k;q)=\left(
\begin{array}{c}
\hat X^{01}_{\sigma\sigma'}(\bm k;q)\\
\hat X^{02}_{\sigma\sigma'}(\bm k;q)\\
\hat X^{03}_{\sigma\sigma'}(\bm k;q)
\end{array}
\right),\\
&&\tilde\Pi_{\sigma\sigma'}(q)=
\left(
\begin{array}{ccc}
\hat\Pi^{11}_{\sigma\sigma'}(q) & \hat\Pi^{12}_{\sigma\sigma'}(q) & \hat\Pi^{13}_{\sigma\sigma'}(q) \\
\hat\Pi^{21}_{\sigma\sigma'}(q) & \hat\Pi^{22}_{\sigma\sigma'}(q) & \hat\Pi^{23}_{\sigma\sigma'}(q) \\
\hat\Pi^{31}_{\sigma\sigma'}(q) & \hat\Pi^{32}_{\sigma\sigma'}(q) & \hat\Pi^{33}_{\sigma\sigma'}(q)
\end{array}
\right).
\end{eqnarray}
Then, Eq.~(\ref{eq.BS_NN3}) can be solved by
\begin{equation}
 \tilde X_{\sigma\sigma'}(\bm k;q)=\left[\tilde I-V\tilde\Pi_{\sigma\sigma'}(q)\right]^{-1}\tilde X^0_{\sigma\sigma'}(\bm k;q).
\end{equation}
The condition for the matrix $\tilde X$ to have poles is given by
\begin{equation}
{\rm det}\left[\tilde I-V\tilde\Pi_{\sigma\sigma'}(q)\right]=0.
\label{eq.pole}
\end{equation}
\par
Figures~\ref{fig.cooperon} (c) and (d) show the energy spectrum of Cooperons obtained by solving Eq.~(\ref{eq.pole}).  Multiple branches of Cooperons appear below the edge of the continuum, because the spin-orbit coupling breaks the rotational symmetry in spin space and lifts the degeneracy between Cooperons with different spin configurations. Figure~\ref{fig.cooperon} (c) illustrates that a bound state of electrons with opposite spins appears in the vicinity of the $K$ point for any $V>0$. The dispersion is symmetric under a rotation of 60 degrees, so the bound state forms also in the vicinity of the $K'$ point. On the other hand, electrons with parallel spins form a bound state in the vicinity of the $\Gamma$ point. This difference between pairs of electrons with parallel and opposite spins can be qualitatively understood by the selection rule in the previous section. Namely, the formation of Cooperons at the $\Gamma$ point corresponds to the inter-valley pairing and at the $K$ and $K'$ points to the intra-valley pairing. 
\par As $V$ is increased, the minima of the dispersions of Cooperons decrease progressively and the condensation of Cooperons with opposite spins first takes place at the $K$ and $K'$ points simultaneously, as shown in Fig.~\ref{fig.cooperon} (d). If $V$ is increased further, Cooperons with parallel spins condense at the $\Gamma$ point.
\par
Figure~\ref{fig.cooperongap} shows the gap of Cooperons with opposite spins at the $K$ point ($\Delta_{\rm VK}$) as well as that of Cooperons with parallel spins at the $\Gamma$ point ($\Delta_{\rm V\Gamma}$) as functions of $V$. $\Delta_{\rm VK}<\Delta_{\rm V\Gamma}$ indeed indicates that the NN interaction favors formation of Cooperons in the vicinity of the $K$ point. The fact that $\Delta_{\rm V K}<\Delta_{\rm U\Gamma}$ for a fixed $t'$ and the critical value $V_{\rm c}$ at the onset of the Cooperon condensation is smaller than $U_{\rm c}$ in Fig.~\ref{fig.cooperongap} also shows that the NN attractive interaction is more effective than the on-site attractive interaction for pair formation.
The condensation of Cooperons with opposite-spin configuration at the $K$ point leads to the spin-triplet intra-valley pairing state, as we will see in the next section.
\par
The two branches within the same spin configuration in Figs.~\ref{fig.cooperon} (c) and (d) correspond to singlet and triplet states of sublattice-pseudospin, whose energy splitting increases as $V$ increases as shown in the figures. In the limit of $t'\to 0$, restoring the SU(2) symmetry in spin space, each of the upper and lower branches becomes doubly degenerate for different spin configurations and there remain two branches of Cooperon bound states. 

\subsection{Berry phase effects}

In this subsection, we illustrate Berry phase effects on the Cooperon condensation at the $K$ and $K'$ points and the intra-valley pairing. For simplicity, we set $t'=0$ and $U=0$.
\par
\begin{figure}
\centering
\includegraphics[width=\linewidth]{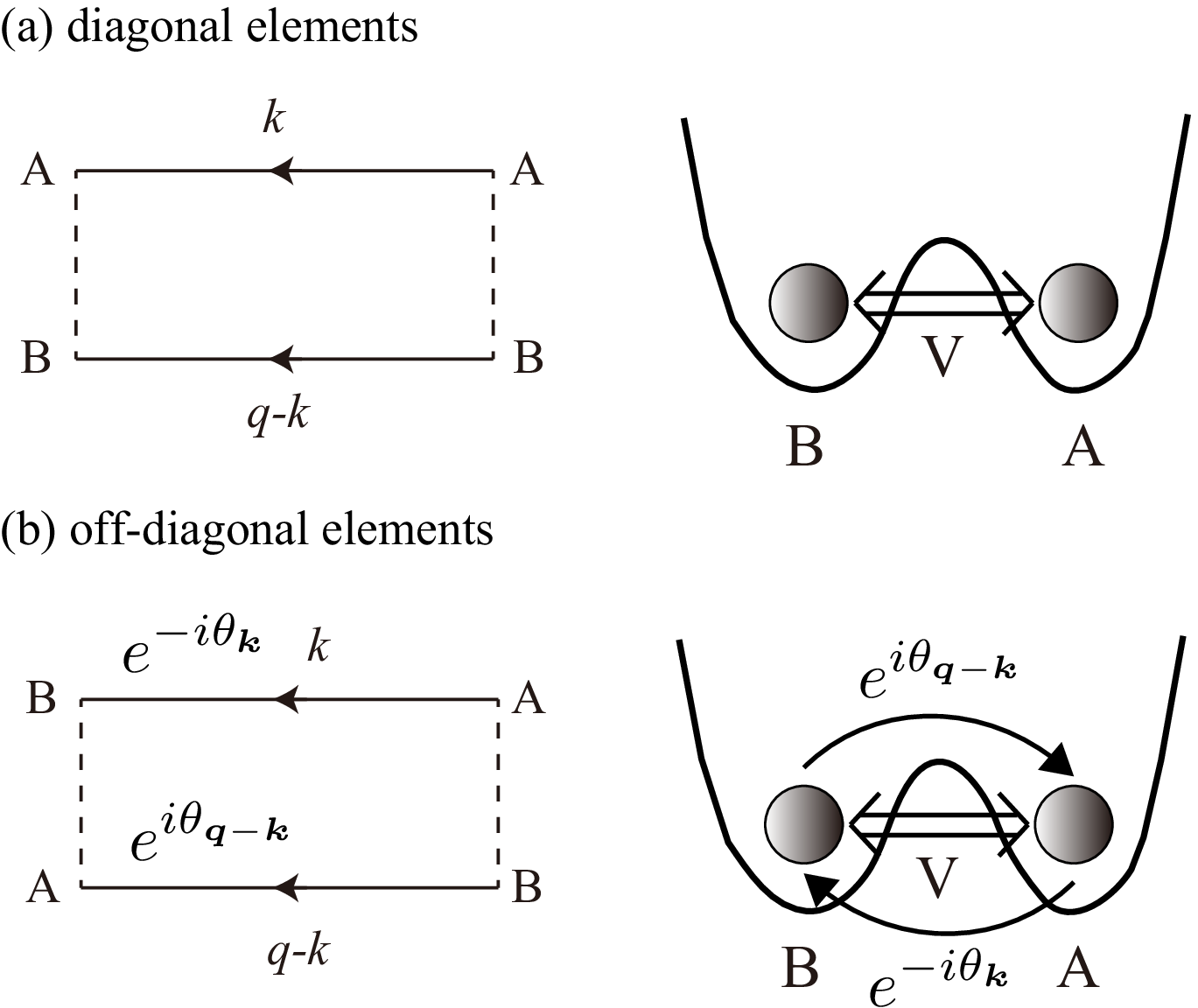}
\caption{{Schematic illustration of the diagonal ((a)) and off-diagonal ((b)) elements of the rung diagram $\hat\pi(\bm k;q)$. The off-diagonal elements involve the phase factors associated with the exchange of electrons in different sublattices.}}
\label{fig.exchange}
\end{figure}

The Green's function, which is independent of electron spin without the SO coupling (${\hat G}_\upa=\hat G_\dna=\hat G$), reads
\beeq
{\hat G}(k)&=&\frac{1}{i\omega_n-\left(
\begin{array}{cc}
0 & \gamma_{\bm k} \\
\gamma_{\bm k}^* & 0
\end{array}
\right)}\nonumber\\
&=&\frac{\frac{1}{2}
\left(
\begin{array}{cc}
1 & e^{i\theta_{\bm k}} \\
e^{-i\theta_{\bm k}} & 1 
\end{array}
\right)
}{i\omega_n-\epsilon_{\bm k}}+
\frac{\frac{1}{2}
\left(
\begin{array}{cc}
1 & -e^{i\theta_{\bm k}} \\
-e^{-i\theta_{\bm k}} & 1 
\end{array}
\right)
}{i\omega_n+\epsilon_{\bm k}}.
\eneq
The diagonal and off-diagonal elements of $\hat\pi(\bm k;q)$ diagrammatically shown in Fig.~\ref{fig.exchange} are given by
\beeq
\pi^{\tau\tau}(\bm k;q)&=&
\frac{1}{4}\frac{2(\epsilon_{\bm k}+\epsilon_{\bm q-\bm k})}{(\epsilon_{\bm k}+\epsilon_{\bm q-\bm k})^2-(i\Omega_n)^2},\\
\pi^{12}(\bm k;q)&=&\frac{e^{i(\theta_{\bm k}-\theta_{\bm q-\bm k})}}{4}\frac{2(\epsilon_{\bm k}+\epsilon_{\bm q-\bm k})}{(\epsilon_{\bm k}+\epsilon_{\bm q-\bm k})^2-(i\Omega_n)^2},
\eneq
where $\pi^{21}=(\pi^{12})^*$. The phase factor of the off-diagonal elements arises from the exchange of electrons in different sublattices as described in Fig.~\ref{fig.exchange}(b).
\par  
We consider {\it intra-valley} pairing and set $\bm q=\bm K$. Assuming the momenta of paired electrons are in the vicinity of the $K'$ point, i.e., $\bm k=\bm K'+\bm p$, $\bm q-\bm k=\bm K'-\bm p$, and by linearizing in the momentum $\bm p$, the phase factor in the off-diagonal elements reduces to
\beeq
e^{i(\theta_{\bm k}-\theta_{\bm q-\bm k})}=e^{i\pi}=-1.
\eneq
The phase factors compensate each other such that the off-diagonal elements of $\hat\Pi^{ij}(q)$ remain finite. This leads to the interference of the direct and exchange processes in Figs.~\ref{fig.exchange} (a) and (b). As a result, the condition of poles (\ref{eq.pole}) with $\bm q=\bm K$ reduces to 
\beeq
\frac{1}{M}\sum_{\bm p}\frac{v_Fp}{4v_F^2p^2-(i\Omega_n)^2}=\frac{1}{6V}.
\eneq
Evaluating the critical value of the interaction strength $V_{\rm c}$ for Cooperon condensation with $\Omega_n=0$, we obtain
\beeq
V_{\rm c}=\frac{8\pi v_F}{3\sqrt{3}p_c},
\eneq
where $p_c$ is a momentum cut-off.
\par
For comparison, we consider now the {\it inter-valley} pairing and set $\bm q=0$. Assuming $\bm k=\bm K'+\bm p$ and linearizing by $\bm p$, the phase factor reduces to
\beeq
e^{i(\theta_{\bm k}-\theta_{\bm q-\bm k})}=e^{2i\phi_{\bm p}},
\eneq
{where $\phi_{\bm p}={\rm arg}(p_x+ip_y)$ is the polar angle of $\bm p$ in the $x-y$ plane.} 
The cancelation of phase factors is absent in this case because of the opposite signs of the Berry phase around $K$ and $K'$. 
The integration over $\bm p$ yields vanishing off-diagonal elements of $\hat\Pi^{ij}(q)$, so the condition (\ref{eq.pole}) with $\bm q=0$ reduces to 
\beeq
\frac{1}{M}\sum_{\bm p}\frac{v_Fp}{4v_F^2p^2-(i\Omega_n)^2}=\frac{1}{3V}.
\eneq
Setting $\Omega_n=0$, we find the interaction strength $V_{\rm c}'$ for Cooperon condensation as
\beeq
V_{\rm c}'=\frac{16\pi v_F}{3\sqrt{3}p_c}=2V_{\rm c}.
\eneq
The critical interaction strength for the onset of the inter-valley pairing is twice as large as that of the intra-valley pairing.
\par
In comparison with the above two cases, we conclude that the interference of the direct and the exchange processes for the intra-valley pairing lowers the energy of Cooperons and yields the Cooperon condensation at $K$ and $K'$. This is consistent with the observation in the previous subsection that the two branches of Cooperon correspond to sublattice-pseudospin singlet and triplet states for $t'=0$,  which arise as an interference effect between the direct and exchange processes in Fig.~\ref{fig.exchange}. Note that the same mechanism indeed works for the Cooperon condensation at $\bm K$ and $\bm K'$ in the case of $t'\neq 0$ due to the phase factors in the off-diagonal elements of the Green's functions in Eqs.~(\ref{eq.Gup}) and (\ref{eq.Gdown}).

\section{Mean-field theory}
\label{sec.mftheory}
In the previous section, we demonstrated that Cooperons composed of electrons with opposite spins condense at $\bm K$ and $\bm K'$, if the NN attractive interaction dominates. This implies the emergence of the intra-valley pairing state in the superconducting phase. In this section, we examine this unconventional superconducting ground state of the KM model with the NN attractive interaction within a mean-field theory. We confirm that the Cooperon condensation at the $K$ and $K'$ points indeed leads to the intra-valley pairing state. 
We use the mean field approach to elucidate some remarkable properties of this state. To simplify the discussion we set $U=0$ and assume only the NN attractive interaction throughout this section.
\par
The NN interaction in momentum space can be written in a standard form~\cite{sigrist-91} as
\begin{eqnarray}
H_{\rm int}&=&\frac{1}{2}\sum_{\bm k,\bm k',\bm q}\sum_{\tau_1\sim\tau_4}
 \sum_{\sigma_1\sim\sigma_4}
 V_{\sigma_1\sigma_2\sigma_3\sigma_4}^{\tau_1\tau_2\tau_3\tau_4}(\bm k,\bm k',\bm q)\nonumber\\
&&\times c_{\bm k\tau_1\sigma_1}^\dagger c_{-\bm k+\bm q\tau_2\sigma_2}^\dagger
 c_{-\bm k'+\bm q\tau_3\sigma_3} c_{\bm k'\tau_4\sigma_4},
\label{eq:factorizedHv}
\end{eqnarray}
where $\bm q$ denotes the center of mass momentum of electron pairs. The matrix element of the interaction reads
\begin{widetext}
\begin{eqnarray}
 V_{\sigma_1\sigma_2\sigma_3\sigma_4}^{\tau_1\tau_2\tau_3\tau_4}(\bm k,\bm k',\bm q)&=& -\frac{V}{2M}\left[(f(\bm k-\bm k')\delta_{\rm ABBA}+f(-\bm k+\bm
			k')\delta_{\rm BAAB})(\delta_{\upa\upa\upa\upa}+\delta_{\upa\dna\dna\upa}+\delta_{\dna\upa\upa\dna}+\delta_{\dna\dna\dna\dna})\right.\nonumber\\
&&\left.-(f(\bm k+\bm k'-\bm q)\delta_{\rm ABAB}+f(-\bm k-\bm
   k'+\bm q)\delta_{\rm BABA})(\delta_{\upa\upa\upa\upa}+\delta_{\upa\dna\upa\dna}+\delta_{\dna\upa\dna\upa}+\delta_{\dna\dna\dna\dna})\right].
\label{eq:matrixV}
\end{eqnarray}
\end{widetext}
Here, we define $\delta_{\sigma_1'\sigma_2'\sigma_3'\sigma_4'}\equiv\delta_{\sigma_1,\sigma_1'}\delta_{\sigma_2,\sigma_2'}\delta_{\sigma_3,\sigma_3'}\delta_{\sigma_4,\sigma_4'}$ and
$\delta_{\tau_1'\tau_2'\tau_3'\tau_4'}\equiv\delta_{\tau_1,\tau_1'}\delta_{\tau_2,\tau_2'}\delta_{\tau_3,\tau_3'}\delta_{\tau_4,\tau_4'}$.
The matrix element satisfies the following relations due to the fermionic anticommutation relations:
\begin{eqnarray}
 &&V_{\sigma_1\sigma_2\sigma_3\sigma_4}^{\tau_1\tau_2\tau_3\tau_4}(\bm k,\bm k', \bm q)= -V_{\sigma_2\sigma_1\sigma_3\sigma_4}^{\tau_2\tau_1\tau_3\tau_4}(-\bm k+\bm q,\bm k',\bm q)\nonumber\\
&&= -V_{\sigma_1\sigma_2\sigma_4\sigma_3}^{\tau_1\tau_2\tau_4\tau_3}(\bm k,-\bm k'+\bm q,\bm q).\label{eq:antisymV}
\end{eqnarray}
\par
The condensation of Cooperons at $K$ and $K'$ with the same interaction strength implies the emergence of two distinct condensates of electron pairs with $\bm q=\bm K$ and $\bm K'$. To describe these condensates, we introduce the two mean-field gap functions with total momenta $\bm K_s$ ($s=\pm$) as
\begin{eqnarray}
\Delta_{\sigma_1\sigma_2}^{\tau_1\tau_2}(\bm k;K_s)&=&\sum_{\bm k'}\sum_{\tau_3,\tau_4}\sum_{\sigma_3,\sigma_4} V_{\sigma_1\sigma_2\sigma_3\sigma_4}^{\tau_1\tau_2\tau_3\tau_4}(\bm k,\bm k', \bm K_s)\nonumber\\
&\times&\langle c_{-{\bm k'+\bm K_s}\tau_3\sigma_3}c_{\bm k'\tau_4\sigma_4}\rangle,
\label{eq:gapfunc}
\end{eqnarray}
where we denote $K_+=K$ and $K_-=K'$. From Eq.~(\ref{eq:antisymV}), the gap functions are antisymmetric with respect to exchange of fermions
\begin{eqnarray}
&&\Delta_{\sigma_1\sigma_2}^{\tau_1\tau_2}(\bm k;K_s)=-\Delta_{\sigma_2\sigma_1}^{\tau_2\tau_1}(-\bm k+\bm K_s;K_s).
\label{eq:antisymDelta}
\end{eqnarray}
We also set the components of the gap functions for equal spins to be zero: $\Delta^{\tau_1\tau_2}_{\sigma\sigma}(\bm k;K_s)=0$, because only Cooperons with opposite spins condense in the presence of the SO coupling. Thus, the non-vanishing components of the gap functions are
\begin{eqnarray}
\Delta^{\rm AB}_{\upa\dna}(\bm k;K_s)&=&\frac{V}{M}\sum_{\bm k'}f(\bm k-\bm k')\langle a_{\bm k'\upa}b_{-\bm k'+\bm K_s\dna}\rangle\nonumber\\
&=&\sum_{j=0}^2\Delta_{j\upa\dna}^s(e_j(\bm k_s)-io_j(\bm k_s))~, \label{eq:DeltaABK}\\
\Delta^{\rm BA}_{\upa\dna}(\bm k;K_s)&=&\frac{V}{M}\sum_{\bm k'}f(\bm k'-\bm k)\langle b_{\bm k'\upa}a_{-\bm k'+\bm K_s\dna}\rangle\nonumber\\
&=&-\sum_{j=0}^2\Delta_{j\dna\upa}^s(e_j(\bm k_s)+io_j(\bm k_s))~, \label{eq:DeltaBAK}
\end{eqnarray}
where $s=\pm$ and $\bm k_{\pm}=\bm k+{\bm K}_{\pm}$. $e_j(\bm k)$ and $o_j(\bm k)$ ($j=0,1,2$) are basis functions within the tight-binding approximation, whose definitions are given in Appendix A.
In deriving Eqs.~(\ref{eq:DeltaABK}) and (\ref{eq:DeltaBAK}), we use the decomposition, 
\begin{eqnarray}
f(\bm k-\bm k') &=&\frac{1}{3}\left\{(e_0(\bm k)-io_0(\bm k))(e_0(\bm k')+io_0(\bm k'))\right.\nonumber\\
&&+(e_1(\bm k)-io_1(\bm k))(e_2(\bm k')+io_2(\bm k')) \\ 
&& \left.+(e_2(\bm k)-io_2(\bm k))(e_1(\bm k')+io_1(\bm k')) \right\}~. \nonumber
\end{eqnarray}
Moreover, $\Delta_{j\sigma_1\sigma_2}^{s}$ ($j=0,1,2$) are coefficients given by
\begin{eqnarray}
\Delta_{0\sigma_1\sigma_2}^{s}&=&\frac{V}{3M}\sum_{\bm k}(e_0(\bm k_s)+io_0(\bm k_s))\langle a_{\bm k\sigma_1}b_{-\bm k+\bm K_s\sigma_2}\rangle,\\
\Delta_{1\sigma_1\sigma_2}^{s}&=&\frac{V}{3M}\sum_{\bm k}(e_2(\bm k_s)+io_2(\bm k_s))\langle a_{\bm k\sigma_1}b_{-\bm k+\bm K_s\sigma_2}\rangle,\\
\Delta_{2\sigma_1\sigma_2}^{s}&=&\frac{V}{3M}\sum_{\bm k}(e_1(\bm k_s)+io_1(\bm k_s))\langle a_{\bm k\sigma_1}b_{-\bm k+\bm K_s\sigma_2}\rangle.
\label{eq:Delta012}
\end{eqnarray}
Other matrix elements can be obtained using the antisymmetric relation~(\ref{eq:antisymDelta}).
\par
Since $e_j(\bm k_s)$ ($o_j(\bm k_s)$) are even (odd) functions of $\bm k_s$, the gap functions in Eqs.~(\ref{eq:DeltaABK}) and (\ref{eq:DeltaBAK}) are in linear combinations of even and odd functions of momentum measured relative to $-\bm K_s$. This is in contrast with the conventional case where the gap function in a spin-singlet (triplet) state is parity-even (odd) with respect to $\bm k\to -\bm k$~\cite{sigrist-91}. 
As first pointed out in Ref.~\cite{uchoa-07}, this parity mixing occurs due to the sublattice degrees of freedom that allows Eq.~(\ref{eq:antisymDelta}) to be satisfied by either even or odd function of $\bm k_s$.
\par
We keep the off-diagonal terms that annihilate and create electron pairs with $\bm q=\bm K$ or $\bm K'$ and neglect other terms in Eq.~(\ref{eq:factorizedHv}) of the type,
\begin{eqnarray}
H_{\rm int}&\simeq& \sum_{s=\pm}H_{K_s},\label{HV_BCS}\\
H_{K_s}&=&\frac{1}{2}\sum_{\bm k,\bm k'}\sum_{\tau_1\sim\tau_4}\sum_{\sigma_1\sim\sigma_4}V_{\sigma_1\sigma_2\sigma_3\sigma_4}^{\tau_1\tau_2\tau_3\tau_4}(\bm k,\bm k',\bm K_s)\nonumber\\
&&\times c_{\bm k\tau_1\sigma_1}^\dagger c_{-\bm k+\bm K_s \tau_2\sigma_2}^\dagger
 c_{-\bm k'+\bm K_s\tau_3\sigma_3} c_{\bm k'\tau_4\sigma_4}.
 \label{eq:HK}
\end{eqnarray}
The interaction term in the mean-field approximation reads
\begin{eqnarray}
H_{K_s}&\simeq& \frac{1}{2}\sum_{\bm k}\sum_{\tau_1,\tau_2}\sum_{\sigma_1,\sigma_2}\left(\Delta_{\sigma_1\sigma_2}^{\tau_1\tau_2}(\bm k;K_s)c_{\bm k\tau_1\sigma_1}^\dagger c_{-\bm k+\bm K_s\tau_2\sigma_2}^\dagger\right.\nonumber\\
&&\left.+{\rm h.c.}\right)+E_{K_s}^{\rm c},\label{eq:Hmf}\\
E_{K_s}^{\rm c}&=&\frac{3M}{V}\sum_{j=0}^2\sum_{\sigma_1\neq\sigma_2}|\Delta^s_{j\sigma_1\sigma_2}|^2.\label{eq:Ec}
\end{eqnarray}
\par
The mean-field Hamiltonian {given in Appendix B} is diagonalized by the Bogoliubov transformation as
\begin{eqnarray}
 H_{\rm MF}=\frac{2}{3}\sum_{\bm k}\sum_{l=1}^3\sum_{\nu} E_{l}^\nu(\bm k)\alpha_{\bm k l\nu}^\dagger\alpha_{\bm k l\nu}+E_g~,
\end{eqnarray}
where $E^\nu_l(\bm k)$ is the quasiparticle spectrum, where $l(=1,2,3)$ denotes the band index and $\nu(=p, h)$ denotes the particle ($p$) or hole ($h$) branch. $\alpha_{\bm k l\nu}$ ($\alpha_{\bm k l\nu}^\dagger$) is the annihilation (creation) operator of a quasiparticle. The ground state energy $E_g$ is thus given by
\begin{eqnarray}
E_g= -\frac{1}{3}\sum_{\bm k}\sum_{l=1}^3\sum_{\nu}E_{l}^\nu(\bm k)+\sum_{s=\pm}E^{\rm c}_{K_s}.
\end{eqnarray}
\par
A symmetry classification of the possible intra-valley pairing states is summarized in Appendix C. To determine the irreducible representation $\Gamma$ of the superconducting ground state, we numerically solve the gap equations~(\ref{eq:DeltaABK}) and (\ref{eq:DeltaBAK}) to evaluate $E_g$ for possible states. We obtain non-vanishing self-consistent solutions for $\Gamma={\rm A}_1$ and B$_1$ which are spin-triplet states with inplane equal-spin pairing. 
The gap function of the A$_1$ state is given by
\begin{eqnarray}
&&\Delta_{\uparrow\downarrow}^{\rm AB}(\bm k;K)=\Delta_{\rm sc}(e_0(\bm k_+)-io_0(\bm k_+))~,\label{eq.gapA1_1}\\
&&\Delta_{\uparrow\downarrow}^{\rm AB}(\bm k;K')=-\Delta_{\rm sc}(e_0(\bm k_-)-io_0(\bm k_-))~,\label{eq.gapA1_2}
\end{eqnarray}
and the B$_1$ state by
\begin{eqnarray}
&&\Delta_{\uparrow\downarrow}^{\rm AB}(\bm k;K)=\Delta_{\rm sc}'(e_0(\bm k_+)-io_0(\bm k_+))~,\\
&&\Delta_{\uparrow\downarrow}^{\rm AB}(\bm k;K')=\Delta_{\rm sc}'(e_0(\bm k_-)-io_0(\bm k_-))~.
\end{eqnarray}
Note that following the Appendix C both spin-triplet states, A$_1$ and B$_1$, are superpositions of sublattice-pseudospin-singlet and -triplet configurations, such that orbital parity can be mixed in these states.
\par
Analogously we develop the mean-field theory and the symmetry classification for the inter-valley pairing states and find that the spin-triplet A$_1$ state has the lowest energy among the possible inter-valley pairing states.
\par
\begin{figure}
\centering
\includegraphics[width=\linewidth]{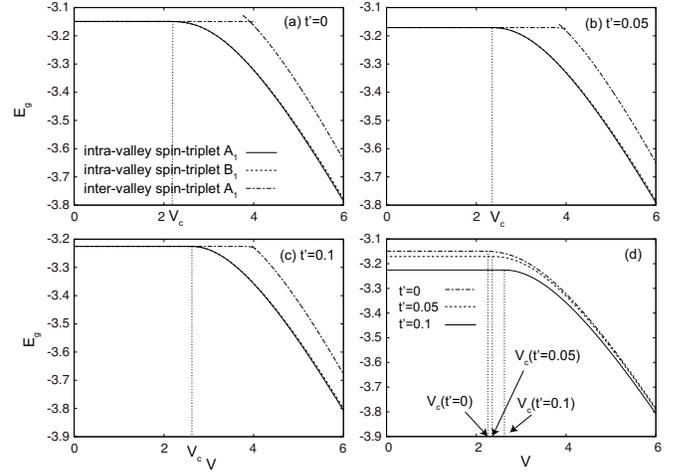}
\caption{Comparison of the energies of the possible superconducting states for a fixed value of $t'$ ((a)$\sim$(c)). Energies are in units of $t$. The solid (dashed) curve shows the energy of the intra-valley-pairing spin-triplet A$_1$ (B$_1$) state, and the dash-dotted curve shows the energy of the inter-valley-pairing spin-triplet A$_1$ state that has the lowest energy among the possible inter-valley pairing states. $V_{\rm c}$ is the strength of the attractive interaction at the onset of the intra-valley-pairing spin-triplet A$_1$ state. Comparison of the energies of the intra-valley-pairing spin-triplet A$_1$ state for different values of $t'$ ((d)).}
\label{fig:gsenergy}
\end{figure}

Fig.~\ref{fig:gsenergy} shows a comparison of the energies of the possible intra and inter-valley pairing states. It demonstrates that the critical strength of $V$ for the onset of the intra-valley pairing states is smaller than that of the inter-valley pairing state, which is consistent with the critical strength of $V$ for Cooperon condensation in Fig.~\ref{fig.cooperongap}. It also shows that the intra-valley pairing states have lower energy than the lowest inter-valley pairing state. The ground state is thus found to be $\Gamma={\rm A}_1$ of the intra-valley spin-triplet pairing state. Note that the ground state is quite close in energy with the intra-valley-pairing spin-triplet B$_1$ state. 
\par
\begin{figure}
\centering
\includegraphics[width=7cm]{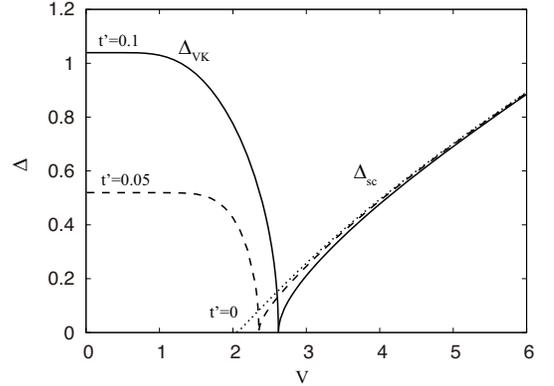}
\caption{Energy gap of a Cooperon at the $K$ and $K'$ points ($\Delta_{\rm VK}$) and the amplitude of the gap function of the intra-valley pairing spin-triplet A$_1$ state ($\Delta_{\rm sc}$) as functions of the strength of the NN attractive interaction $V$. Energies are in units of $t$.}
\label{fig:gapfig}
\end{figure}

Fig.~\ref{fig:gapfig} shows a plot of the amplitude $\Delta_{\rm SC}$ of the gap function in Eqs.~(\ref{eq.gapA1_1}) and (\ref{eq.gapA1_2}), which we take positive and real without loss of generality, together with the energy gap of a Cooperon at the $K$ and $K'$ points $\Delta_{\rm VK}$ as functions of $V$. Note that the interaction strength for the Cooperon condensation precisely matches with the onset of the superconducting phase. The consistency of the two independent unbiased approaches shows that the condensation of Cooperons with antiparallel spins at the $K$ and $K'$ points leads to the intra-valley spin-triplet pairing state.
\begin{figure}
\centering
\includegraphics[width=7cm]{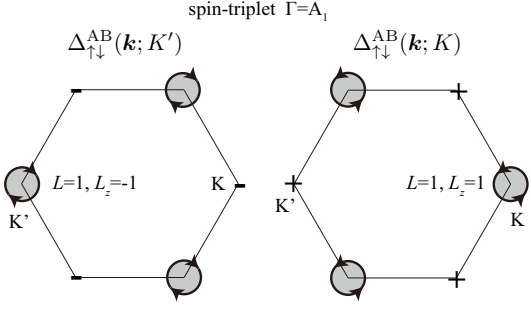}
\caption{Schematic illustration of angular momentum of Cooper pairs in the intra-valley-pairing spin-triplet A$_1$ state. The symbols ``$+$'' and ``$-$'' denote the signs of the $s$-wave component of the gap function in the vicinity of $K$ or $K'$.}
\label{fig:tripletA1}
\end{figure}
\par
The expansions of the gap functions in the vicinity of $K$ and $K'$ read
\begin{eqnarray}
\Delta_{\uparrow\downarrow}^{\rm AB}(\bm k;K)&\simeq&
\left\{
\begin{array}{ll}
\displaystyle
\dfrac{\sqrt{3}}{2}\Delta_{\rm sc}(p_x+ip_y), & (\bm k\simeq \bm K),\\[2mm]
3\Delta_{\rm sc}, & (\bm k\simeq \bm K'),
\end{array}
\right.\\\nonumber\\
\Delta_{\uparrow\downarrow}^{\rm AB}(\bm k;K')&\simeq&
\left\{
\begin{array}{ll}
-3\Delta_{\rm sc}, & (\bm k\simeq \bm K),\\[2mm]
 \dfrac{\sqrt{3}}{2}\Delta_{\rm sc}(p'_x-ip'_y), & (\bm k\simeq \bm K'),
 \end{array}
 \right.
 \label{eq:gapexpansion}
\end{eqnarray}
where $\bm p=\bm k-\bm K$ and $\bm p'=\bm k-\bm K'$. The above expansion shows that $\Delta_{\uparrow\downarrow}^{\rm AB}(\bm k;K_\pm)$ has a point node at $K_\pm$ due to the dominant $p$-wave component and a Cooper pair with center of mass momentum $\bm q=\bm K$ ($\bm K'$) has angular momentum along the $z$-axis, $L_z=1$ (-1), which is schematically illustrated in Fig.~\ref{fig:tripletA1}. The superconducting state has time-reversal symmetry, because the total angular momentum of the system is zero. The fact that the pairing state has opposite chirality for $K$ and $K'$ identifies this state a ``helical'' valley-triplet state, the valley-analog to the $^3$He-B phase in 2D~\cite{leggett-75, qi-09}.
\par
\begin{figure}
\centering
\includegraphics[width=\linewidth]{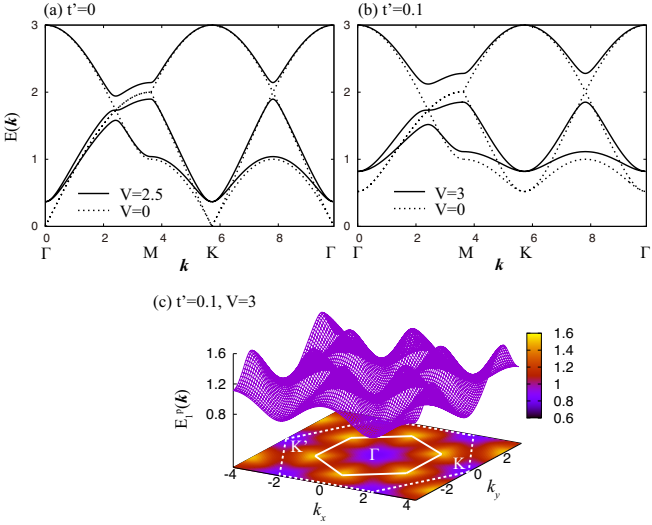}
\caption{(Color online) Quasiparticle band structure in the intra-valley-pairing spin-triplet A$_1$ state when $t'=0$ ((a)) and $t'=0.1$ ((b)). Energies are in units of $t$. The solid (dotted) curves in (a) and (b) are the energy spectrum in the superconducting (normal) state. The panel (c) shows the 3D plot of the lowest band in (b). The solid line in the contour plot in (c) represents the BZ of the quasiparticle band in the superconducting state, while the dashed line that of the original lattice structure.}
\label{fig:qpspectrum}
\end{figure}
{Figures~\ref{fig:qpspectrum} (a)$\sim$(c) show the quasiparticle spectrum in the intra-valley-pairing spin-triplet A$_1$ state. In the contour plot in Fig.~\ref{fig:qpspectrum} (c), the original Brillouin zone (BZ) of the honeycomb lattice is folded into one third so that the $\Gamma$, $K$, and $K'$ points are identical in the superconducting state. As a result, the quasiparticle band in the reduced BZ splits into the three bands and energy gap opens between them as shown in Figs.~\ref{fig:qpspectrum} (a) and (b). Note that the degeneracy of the lowest two bands is not lifted at $\Gamma$, which is time-reversal invariant point in the BZ, due to Kramers' theorem. The folding of the BZ implies the emergence of the spatial pattern of the pair amplitude that spontaneously breaks the translational symmetry of the original lattice structure in the superconducting phase, as we will discuss in the next subsection. Figure~\ref{fig:qpspectrum} (a) shows that Dirac fermions are gapped at $\Gamma$ when $t'=0$ due to the $s$-wave component of the gap function in Eq.~(\ref{eq:gapexpansion}). In Fig.~\ref{fig:qpspectrum} (b) when $t'\neq 0$, the energy gap at $\Gamma$ gets larger in the superconducting phase.}

{
\subsection{Spatial modulation of pair amplitude}
The pair amplitude in NN bonds reads
\begin{eqnarray}
&&\langle a_{i\sigma}b_{j\bar\sigma}\rangle=\chi_{\sigma\bar\sigma\alpha}^+e^{i\bm K\cdot{\bm r}_i}+\chi_{\sigma\bar\sigma\alpha}^-e^{i\bm K'\cdot{\bm r}_i},\nonumber\\
&&\chi_{\sigma\bar\sigma\alpha}^\pm=\frac{1}{M}\sum_{\bm k}e^{i\bm k_\pm\cdot{\bm \delta}_\alpha}\langle a_{\bm k\sigma}b_{-\bm k_\pm\bar\sigma}\rangle,
\end{eqnarray}
where $\bm r_j=\bm r_i+{\bm \delta}_\alpha$. For each direction of the NN bonds, the pair amplitude for the intra-valley-pairing spin-triplet A$_1$ state can be calculated as
\beeq
&&\langle a_{i\upa}b_{j\dna}\rangle=\langle a_{i\dna}b_{j\upa}\rangle\nonumber\\
&&\propto\left\{
\begin{array}{lc}
-\sin(\bm K\cdot{\bm r}_i), & (\alpha=1),\\[2mm]
\sin\big(\bm K\cdot{\bm r}_i+\dfrac{\pi}{3}\big), & (\alpha=2),\\[2mm]
\sin\big(\bm K\cdot{\bm r}_i-\dfrac{\pi}{3}\big), & (\alpha=3).
\end{array}
\right.
\label{eq:pairampA1}
\eneq
The above equation indeed demonstrates that the intra-valley pairing state is a PDW state and it is analogous to the Larkin-Ovchinikov state~\cite{larkin-64} in which the amplitude of the gap function spatially modulates. Since the pair amplitude in Eq.~(\ref{eq:pairampA1}) does not involve phase modulation, the ground state has no local supercurrents.}
\begin{figure}
\centering
\includegraphics[width=7cm]{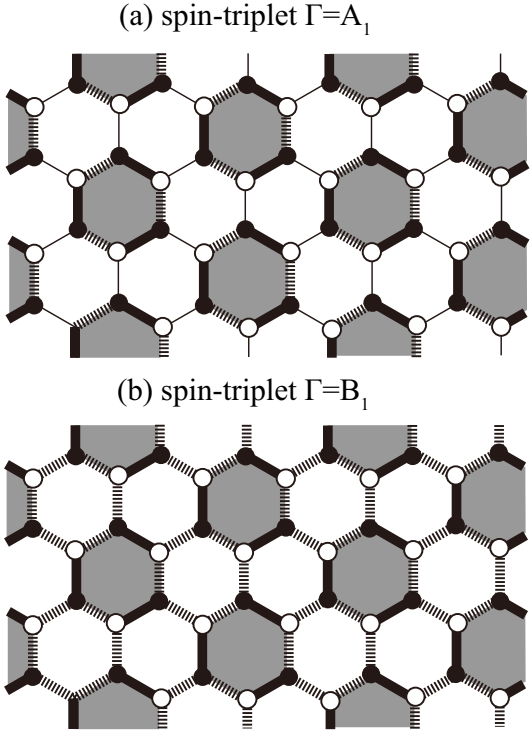}
\caption{Spatial modulation of the pair amplitude in the intra-valley-pairing spin-triplet A$_1$ ((a)) and B$_1$ ((b)) states. The thick solid (dashed) lines represent positive (negative) pair amplitude, while the thin lines do nodes of the pair amplitude. The gray regions highlight the hexagons on which the pair amplitude alternates its sign on the adjacent bonds. Note that the symmetry of the honeycomb lattice $C_{6v}$ is lowered to $C_{3v}$ in the Kekul\'e patterns in (a) and (b).}
\label{fig:pdw}
\end{figure}
\par
{Figure~\ref{fig:pdw} (a) shows the spatial modulation of the pair amplitude in Eq.~(\ref{eq:pairampA1}). It is remarkable that the pair amplitude spontaneously breaks the translational symmetry of the honeycomb lattice and exhibits a {\it Kekul\'e pattern}: The honeycomb lattice consists of the linked hexagons on which the pair amplitude alternates its sign on the adjacent bonds and the pair amplitude has nodes on the bonds that connect these hexagons. The superconducting state with the Kekul\'e pattern in Fig.~\ref{fig:pdw} (a) was recently proposed in the context of superconductivity in graphene due to NN attractive interaction in Ref.~\cite{roy-10} and referred to as the {\it p}-Kekul\'e state. This exotic superconducting state recently attracts attention in the study of graphene \cite{kunst-15,faye-16}. Ref.~\cite{roy-10} predicted the phase transition from the semimetallic phase into the $p$-Kekul\'e state in graphene. Thus, our present mean-field analysis is consistent with that in Ref.~\cite{roy-10} based on the variational ansatz in the case of $t'=0$.}
\par
{The pair amplitude of the spin-triplet intra-valley-pairing B$_1$ state that is competing with the ground state can be calculated as
\beeq
&&\langle a_{i\upa}b_{j\dna}\rangle=\langle a_{i\dna}b_{j\upa}\rangle\nonumber\\
&&\propto\left\{
\begin{array}{lc}
\cos(\bm K\cdot{\bm r}_i), & (\alpha=1),\\[2mm]
\cos\big(\bm K\cdot{\bm r}_i+\dfrac{2\pi}{3}\big), & (\alpha=2),\\[2mm]
\cos\big(\bm K\cdot{\bm r}_i-\dfrac{2\pi}{3}\big), & (\alpha=3).
\end{array}
\right.
\label{eq:pairampB1}
\eneq
Figure~\ref{fig:pdw} (b) shows the spatial modulation of the pair amplitude in Eq.~(\ref{eq:pairampB1}). It exhibits another Kekul\'e pattern where the pair amplitude is negative on the bonds that connect the hexagons on which the pair amplitude alternates its sign on the adjacent bonds and it is called $s$-Kekul\'e state in Ref.~\cite{roy-10}. However, although a discontinuous transition between the $s$-Kekul\'e and $p$-Kekul\'e states in the superconducting phase is predicted in Ref.~\cite{roy-10}, we do not find such a transition when $t'=0$.}

\section{Conclusions}
\label{sec.conclusions}

{To summarize, we investigated the possibility of the intra-valley pairing state in the KM model with short-ranged attractive interaction. We found that the NN attractive interaction induces Cooperon condensation at $K$ and $K'$ and leads to the emergence of the intra-valley-pairing spin-triplet superconducting state with $\Gamma={\rm A}_1$ of the point group $C_{6v}$. We found that the pair amplitude spontaneously breaks the translational symmetry and exhibit a $p$-Kekul\'e pattern in this exotic PDW superconducting state. 
As a ``valley-helical" state it is a topological superconducting phase. 
Although we restricted our analysis to half-filling, the intra-valley-pairing spin-triplet A$_1$ state can be indeed realized, if the system is lightly doped from half-filling.}
\par
{Since the on-site interaction is repulsive in real materials due to the Coulomb interaction, the NN attractive interaction may play a dominant role for the superconductivity in Li-decorated monolayer graphene and TMDs. Our prediction could be confirmed in these systems by observing the $p$-Kekul\'e patter in Fig.~\ref{fig:pdw} that is a clear signature of the intra-valley-pairing spin-triplet A$_1$ state. For instance, it would be interesting to observe the Kekul\'e pattern by a probe that has resolution in atomic-scale such as scanning tunneling microscope \cite{gutierrez-16}.}

\section*{Acknowledgments}
S. T. thanks I. Danshita, S. Fujimoto, S. Konabe, Y. Ohashi, M. Sato, K. Totsuka, D. Yamamoto, and Y. Yanase for useful comments and discussions. This work was supported by JSPS KAKENHI Grant Number 15H05885 (J-Physics). S. T. was supported by Grant-in-Aid for Scientific Research, No. 24740276. S.T., J.G., and E.A. are grateful to the Pauli Center for Theoretical Physics of ETH Zurich for hospitality.

\appendix

\section{Basis functions}

The basis functions within the tight-binding approximation are given by
\begin{eqnarray}
&&e_0(\bm k)=\sum_{n=1}^3 \cos\bm k\cdot\bm\delta_n,\quad o_0(\bm k)=\sum_{n=1}^3 \sin\bm k\cdot\bm\delta_n,\\
&&e_1(\bm k)=\sum_{n=1}^3 z^{n-1}\cos\bm k\cdot\bm\delta_n,\quad e_2(\bm k)=e_1(\bm k)^*,\\
&&o_1(\bm k)=\sum_{n=1}^3 z^{n-1}\sin\bm k\cdot\bm\delta_n,\quad o_2(\bm k)=o_1(\bm k)^*,
\end{eqnarray}
where we denote $z = e^{i 2\pi/3}$. 

\section{Mean-field Hamiltonian}
The mean-field Hamiltonian is given by
\begin{eqnarray}
H_{\rm MF}&=&=\frac{1}{3}\sum_{\bm k}\Psi^\dagger(\bm k) \hat{\mathcal H}(\bm k)\Psi(\bm k),\\
\Psi(\bm k)&=&(\psi_{\bm k\upa},\psi_{\bm k_+\upa},\psi_{\bm k_-\upa},\psi_{-\bm k\dna}^\dagger,\psi_{-\bm k_+\dna}^\dagger,\psi_{-\bm k_-\dna}^\dagger )^T,\\
\hat{\mathcal H}(\bm k)&=&
\left(
\begin{array}{cc}
\hat{\mathcal H}_{\upa\upa}(\bm k) & \hat{\mathcal H}_{\upa\dna}(\bm k) \\
\hat{\mathcal H}_{\dna\upa}(\bm k) & \hat{\mathcal H}_{\dna\dna}(\bm k)
\end{array}
\right),
\label{eq:Hmf}
\end{eqnarray}
where $\bm k_\pm=\bm k\pm \bm K$. $\hat{\mathcal H}_{\sigma_1 \sigma_2}(\bm k)$ ($\sigma_1,\sigma_2=\upa,\dna$) are the 6$\times$6 matrices defined as 
\begin{eqnarray}
 \hat{\mathcal H}_{\upa\upa}(\bm k)&=&
\left(
\begin{array}{ccc}
\hat\xi_{\bm k} & 0& 0\\
0& \hat\xi_{\bm k_+}& 0\\
0 &0 & \hat\xi_{\bm k_-}
\end{array}
\right),\label{eq:Huu} \\
\hat{\mathcal H}_{\dna\dna}(\bm k)&=&
\left(
\begin{array}{ccc}
-\hat\xi_{\bm k} & 0& 0\\
0 & -\hat\xi_{\bm k_+} & 0\\
0  &0 & -\hat\xi_{\bm k_-}
\end{array}
\right),\label{eq:Hdd}\\
 \hat{\mathcal H}_{\upa\dna}(\bm k)&=&
\left(
\begin{array}{ccc}
0 & \hat\Delta(\bm k;K_-) &  \hat\Delta(\bm k;K_+) \\
\hat\Delta(\bm k_+;K_+) & 0 & \hat\Delta(\bm k_+;K_-) \\
\hat\Delta(\bm k_-;K_-) & \hat\Delta(\bm k_-;K_+) & 0
\end{array}
\right),\label{eq:Hud} \\
 \hat{\mathcal H}_{\dna\upa}(\bm k)&=&(\hat{\mathcal H}_{\upa\dna}(\bm k))^\dagger.\label{eq:Hdu}
\end{eqnarray}
Here, we defined
\begin{eqnarray}
\hat\xi_{\bm k}&=&
\left(
\begin{array}{cc}
\zeta_{\bm k}-\mu & \gamma_{\bm k}\\
\gamma_{\bm k}^* & -\zeta_{\bm k}-\mu
\end{array}
\right),\\
\hat\Delta(\bm k;K_s)&=&
\left(
 \begin{array}{cc}
 & \Delta_{\upa\dna}^{AB}(\bm k;K_s)\\
\Delta_{\upa\dna}^{BA}(\bm k;K_s) & 
\end{array}
\right).\label{eq:gapmatrix}
\end{eqnarray}

\section{Symmetry classification of intra-valley pairing states}
\label{appendixA}
We present a symmetry classification of intra-valley pairing states according to the irreducible representations of the point group of the honeycomb lattice $C_{6v}$. 
The intra-valley-paring states, i.e., valley-pseudospin-triplet states can be classified into (1) spin-singlet, sublattice-pseudospin-singlet state, (2) spin-singlet, sublattice-pseudospin-triplet state, (3) spin-triplet, sublattice-pseudospin-singlet, and (4) spin-triplet, sublattice-pseudospin-triplet states. The gap function for each of (1)$\sim$(4) can be written as
\begin{eqnarray}
&&\Delta_{\sigma_1\sigma_2}^{\tau_1\tau_2,s_1s_2}(\bm k)=\nonumber\\
&&\left\{
\begin{array}{cl}
(1) & \psi^\nu(\bm k)(i\sigma_y)_{\sigma_1\sigma_2}(i\tau_y)^{\tau_1\tau_2}(is_\nu s_y)^{s_1s_2}, \\[2mm]
(2) & \psi^{\nu\rho}(\bm k) (i\sigma_y)_{\sigma_1\sigma_2}(i\tau_\nu\tau_y)^{\tau_1\tau_2}(is_\rho s_y)^{s_1s_2}, \\[2mm]
(3) & d_\nu^\rho(\bm k)(i\sigma_\nu \sigma_y)_{\sigma_1\sigma_2}(i\tau_y)^{\tau_1\tau_2}(is_\rho s_y)^{s_1s_2}, \\[2mm]
(4) & d_\nu^{\rho\eta}(\bm k)(i\sigma_\nu \sigma_y)_{\sigma_1\sigma_2}(i\tau_\rho\tau_y)^{\tau_1\tau_2}(is_\eta s_y)^{s_1s_2},
\end{array}
\right.
\label{eq:generalform}
\end{eqnarray}
where $\sigma_\rho$, $\tau_\rho$, and $s_\rho$ ($\rho=x,y,z$) are the Pauli matrices of spin, sublattice-pseudospin, and valley-pseudospin, respectively.
Note that $\sigma_i=\uparrow,\downarrow$, $\tau_i={\rm A,B}$, and $s_i=\pm$ ($i=1,2$). In Eq.~(\ref{eq:generalform}), summation is taken over repeated indices.
If electrons in the vicinity of $K$ ($K'$) form a pair, the center of mass momentum of the pair is $\bm K'$ ($\bm K$). In Sec.~\ref{sec.mftheory}, we thus denote 
\begin{eqnarray}
\Delta_{\sigma_1\sigma_2}^{\tau_1\tau_2,++}(\bm k)&=&\Delta_{\sigma_1\sigma_2}^{\tau_1\tau_2}(\bm k;K'),\\
\Delta_{\sigma_1\sigma_2}^{\tau_1\tau_2,--}(\bm k)&=&\Delta_{\sigma_1\sigma_2}^{\tau_1\tau_2}(\bm k;K)
\end{eqnarray}
\par
Eq.~(\ref{eq:antisymDelta}) restricts the parity of the order parameters as
\begin{eqnarray}
\begin{array}{ll}
\psi^+(-\bm k+\bm K)=-\psi^+(\bm k), & \psi^-(-\bm k+\bm K')=-\psi^-(\bm k), \\[2mm]
\psi^{\nu +}(-\bm k+\bm K)=\psi^{\nu +}(\bm k), & \psi^{\nu -}(-\bm k+\bm K')=\psi^{\nu -}(\bm k), \\[2mm]
d^+_\nu(-\bm k+\bm K)=d^+_\nu(\bm k), & d^-_\nu(-\bm k+\bm K')=d^-_\nu(\bm k), \\[2mm]
d^{\rho +}_\nu(-\bm k+\bm K)=-d^{\rho +}_\nu(\bm k), & d^{\rho -}_\nu(-\bm k+\bm K')=-d^{\rho -}_\nu(\bm k),
\end{array}
\end{eqnarray}
where we denote $\psi^\pm\equiv \psi^x\pm i \psi^y$ and $d^\pm\equiv d^x\pm i d^y$ in valley-pseudospin space.
\par
In Table \ref{classification-gap}, we list the basis gap functions that satisfy the above restriction for parity. 
Bold symbols are vectors in spin space, whereas symbols with arrow and tilde denote vectors in sublattice-pseudospin and valley-pseudospin spaces, respectively. We denote $\tilde x_{\pm}=\tilde x\pm i\tilde y$.
Note that in Table \ref{classification-gap}, we assume inter-sublattice pairing of electrons with opposite spins due to the NN attractive interaction and the SO interaction. Therefore, the $d$-vector of the sublattice-pseudospin-triplet state is parallel to $\vec z$ ($\vec\psi, \vec{\bm d}\parallel \vec z$) and that of spin-triplet state is parallel to $\bm z$ ($\bm d\parallel \bm z$).
\begin{table*}
\caption{Basis gap functions of intra-valley pairing states for each of the irreducible representations $\Gamma$ of the point group $C_{6v}$. $\sigma_\rho$, $\tau_\rho$, and $s_{\rho}$ ($\rho=x,y,z$) are the Pauli matrices of spin, sublattice-pseudospin, and valley-pseudospin, respectively.}
\vspace{1cm}
(a) spin-singlet states
\begin{ruledtabular}
\begin{tabular}{l || l l}
$\Gamma$ $\backslash$ $\Delta_{\sigma_1\sigma_2}^{\tau_1\tau_2,s_1s_2}(\bm k)$ & sublattice-pseudospin-singlet & sublattice-pseudospin-triplet \\[2mm]
 & $\psi^\nu(\bm k)(i\sigma_y)_{\sigma_1\sigma_2}(i\tau_y)^{\tau_1\tau_2}(is_\nu s_y)^{s_1s_2}$ & $\psi^{\nu\rho}(\bm k) (i\sigma_y)_{\sigma_1\sigma_2}(i\tau_\nu\tau_y)^{\tau_1\tau_2}(is_\rho s_y)^{s_1s_2}$ \\[2mm]
\hline
A$_1$ & $\tilde\psi_{\rm A_1}(\bm k)=o_0(\bm k_-)\tilde{x}_+-o_0(\bm k_+)\tilde{x}_-$ & $\tilde{\vec\psi}_{\rm A_1}(\bm k)=(e_0(\bm k_-)\tilde x_+ - e_0(\bm k_+)\tilde x_-)\otimes\vec z$ \\
\\
B$_1$&$\tilde\psi_{\rm B_1}(\bm k)=o_0(\bm k_-)\tilde{x}_++o_0(\bm k_+)\tilde{x}_-$ & $\tilde{\vec \psi}_{\rm B_1}(\bm k)=(e_0(\bm k_-)\tilde x_+ + e_0(\bm k_+)\tilde x_-)\otimes\vec z$ \\
\\
E$_1$ & $\tilde\psi_{({\rm E}_1,1)}(\bm k)=o_2(\bm k_-)\tilde x_++o_2(\bm k_+)\tilde x_-$ & $\tilde{\vec \psi}_{({\rm E}_1,1)}(\bm k)=(e_2(\bm k_-)\tilde x_++e_2(\bm k_+)\tilde x_-)\otimes\vec z$ \\
& $\tilde\psi_{({\rm E}_1,2)}(\bm k)=o_1(\bm k_-)\tilde x_++o_1(\bm k_+)\tilde x_-$ & $\tilde{\vec\psi}_{({\rm E}_1,2)}(\bm k)=(e_1(\bm k_-)\tilde x_++e_1(\bm k_+)\tilde x_-)\otimes\vec z$ \\
\\
E$_2$ & $\tilde\psi_{({\rm E}_2,1)}(\bm k)=o_2(\bm k_-)\tilde x_+-o_2(\bm k_+)\tilde x_-$ & $\tilde{\vec\psi}_{({\rm E}_2,1)}(\bm k)=(e_2(\bm k_-)\tilde x_+ - e_2(\bm k_+)\tilde x_-)\otimes\vec z$\\
& $\tilde\psi_{({\rm E}_2,2)}(\bm k)=o_1(\bm k_-)\tilde x_+-o_1(\bm k_+)\tilde x_-$ & $\tilde{\vec \psi}_{({\rm E}_2,2)}(\bm k)=(e_1(\bm k_-)\tilde x_+ - e_1(\bm k_+)\tilde x_-)\otimes\vec z$
\end{tabular}
\end{ruledtabular}
\vspace{1cm}
(b) spin-triplet states
\begin{ruledtabular}
\begin{tabular}{l || l l}
$\Gamma$ $\backslash$ $\Delta_{\sigma_1\sigma_2}^{\tau_1\tau_2,s_1s_2}(\bm k)$ & sublattice-pseudospin-singlet & sublattice-pseudospin-triplet \\[2mm]
 & $d_\nu^\rho(\bm k)(i\sigma_\nu \sigma_y)_{\sigma_1\sigma_2}(i\tau_y)^{\tau_1\tau_2}(is_\rho s_y)^{s_1s_2}$ & $d_\nu^{\rho\eta}(\bm k)(i\sigma_\nu \sigma_y)_{\sigma_1\sigma_2}(i\tau_\rho\tau_y)^{\tau_1\tau_2}(is_\eta s_y)^{s_1s_2}$ \\[2mm]
\hline
A$_1$ & $\tilde{\bm d}_{{\rm A}_1}(\bm k)=(e_0(\bm k_-) \tilde x_+ + e_0(\bm k_+)\tilde x_-)\otimes\bm z$ & $\tilde{\vec{\bm d}}_{{\rm A}_1}(\bm k)=(o_0(\bm k_-) \tilde x_+ + o_0(\bm k_+)\tilde x_-)\otimes\bm z\otimes\vec z$ \\
\\
B$_1$ & $\tilde{\bm d}_{{\rm B}_1}(\bm k)=(e_0(\bm k_-) \tilde x_+ - e_0(\bm k_+) \tilde x_-)\otimes\bm z$  & $\tilde{\vec{\bm d}}_{{\rm B}_1}(\bm k)=(o_0(\bm k_-) \tilde x_+ - o_0(\bm k_+) \tilde x_-)\otimes\bm z\otimes\vec z$ \\
\\
E$_1$ & $\tilde{\bm d}_{({\rm E}_1,1)}(\bm k)=(e_2(\bm k_-) \tilde x_+ - e_2(\bm k_+) \tilde x_-)\otimes\bm z$ & $\tilde{\vec{\bm d}}_{({\rm E}_1,1)}(\bm k)=(o_2(\bm k_-) \tilde x_+ - o_2(\bm k_+) \tilde x_-)\otimes\bm z\otimes\vec z$ \\
& $\tilde{\bm d}_{({\rm E}_1,2)}(\bm k)=(e_1(\bm k_-) \tilde x_+ - e_1(\bm k_+) \tilde x_-)\otimes\bm z$ & $\tilde{\vec{\bm d}}_{({\rm E}_1,1)}(\bm k)=(o_1(\bm k_-) \tilde x_+ - o_1(\bm k_+) \tilde x_-)\otimes\bm z\otimes\vec z$ \\
\\
E$_2$ & $\tilde{\bm d}_{({\rm E}_2,1)}(\bm k)=(e_2(\bm k_-) \tilde x_+ + e_2(\bm k_+) \tilde x_-)\otimes\bm z$ & $\tilde{\vec{\bm d}}_{({\rm E}_2,1)}(\bm k)=(o_2(\bm k_-) \tilde x_+ + o_2{\bm k} \tilde x_-)\otimes\bm z\otimes\vec z$ \\
& $\tilde{\bm d}_{({\rm E}_2,1)}(\bm k)=(e_1(\bm k_-) \tilde x_+ + e_1(\bm k_+) \tilde x_-)\otimes\bm z$ & $\tilde{\vec{\bm d}}_{({\rm E}_2,2)}(\bm k)=(o_1(\bm k_-) \tilde x_+ + o_1(\bm k_+) \tilde x_-)\otimes\bm z\otimes\vec z$
\end{tabular}
\end{ruledtabular}
\label{classification-gap}
\end{table*}
\par
In Table~\ref{classification-gap}, both the sublattice-singlet and triplet states have the basis gap functions in the same irreducible representation. 
In general, mixing of basis functions is possible if they are in the same irreducible representation. In fact, pairing with the NN attractive interaction induces the mixing of sublattice-pseudospin-singlet and -triplet states. 
The general form of the gap function of the mixed states are given by 
\begin{eqnarray}
&&\Delta_{\sigma_1\sigma_2}^{\tau_1\tau_2,s_1s_2}(\bm k)=\nonumber\\
&&\left\{
\begin{array}{l}
({\rm spin-singlet})\\[2mm]
\left[\psi^\nu(\bm k)\tau_y+\psi^{z\nu}(\bm k)\tau_x\right]^{\tau_1\tau_2} (i\sigma_y)_{\sigma_1\sigma_2}(is_\nu s_y)^{s_1s_2},\\[3mm]
(\rm spin-triplet)\\[2mm]
\left[d_{z}^\nu(\bm k)(i\tau_y)-id_{z}^{z\nu}(\bm k)\tau_x\right]^{\tau_1\tau_2} (\sigma_x)_{\sigma_1\sigma_2}(is_\nu s_y)^{s_1s_2},
\end{array}
\right.
\label{eq:mixing}
\end{eqnarray}
In Eq.~(\ref{eq:mixing}), the ratio of the mixing is fixed by the form of the NN interaction. On the other hand, mixing of spin-singlet and spin-triplet does not occur, though they have the basis gap functions in the same irreducible representations.

\end{document}